\documentclass[]{raa}
\usepackage{deluxetable}
\usepackage{amssymb}            
\usepackage{graphicx,times}
\usepackage{natbib}
\usepackage{color}

\providecommand{\bm}[1]{\mbox{\bol
dmath$#1$\unboldmath}}

\begin{document}

\title{A Pilot Study of Catching High-$z$ GRBs and Exploring Circumburst Environment  in  the Forthcoming \it SVOM \bf Era}

\volnopage{}
	\setcounter{page}{1}

	\author{J. Wang	
	\inst{1,2}
	\and Y. L. Qiu\inst{2}
        \and J. Y. Wei\inst{2,3}
	}
 \institute{Guangxi Key Laboratory for Relativistic Astrophysics, School of Physical Science and Technology, Guangxi University, Nanning 530004, China; {wj@nao.cas.cn} 
  \and Key Laboratory of Space Astronomy and Technology, National Astronomical Observatories,
Chinese Academy of Sciences, Beijing 100012, China
\and School of Astronomy and Space Science, University of Chinese Academy of Sciences, Beijing, China\\
%
}

\abstract{
Rapid spectroscopy of GRB afterglows is an important and hard task. 
Based on the archival XS-GRB spectral database, we here perform a
pilot study on the afterglow's spectroscopy in the forthcoming \it SVOM \rm era in two aspects.
At first,  our simulation indicates that the
color acquired from the \it SVOM\rm/VT blue and red channels is effective in discriminating between low-$z$ ($z\lesssim3$) and distant GRB candidates until $z\sim6$.
Secondly,  by doubling the sample size, we find that the  previously proposed global photoionization response of the circumburst gas to the prompt emission (i.e., the CIV/CII-$L_{\mathrm{iso}}/E^2_{\mathrm{peak}}$ relationship) is 
roughly confirmed, although the confirmation is dissatisfactory at the
large $L_{\mathrm{iso}}/E^2_{\mathrm{peak}}$ end. We believe that
this issue can be further addressed in the \it SVOM \rm era by larger spectroscopy sample, 
thanks to its capability of rapid identification of optical candidates of afterglow and its anti-solar pointing strategy.
}

 \authorrunning{Wang et al. }            
 \titlerunning{Catching High-$z$ GRBs by \it SVOM}  
 \maketitle

\section{Introduction}

Long gamma-ray bursts (LGRBs), which are believed to originate from the core-collapse
of young massive stars ($\geq25M_\odot$; e.g., Woosley \& Bloom 2006; Hjorth \& Bloom 2012 and references therein),
are the most powerful explosions occurring in the universe.
Because of their extreme high luminosities,  GRBs can be detected and studied from local up to high redshifts
(e.g., Salvaterra et al. 2009; Tanvir et al. 2009). So far, GRB\,090429B with a  photometric redshift of
$z=9.4$ (Cucchiara et al. 2011) is the most distant one reported in the literature.
Up to date, there are only 9 \it Swift \rm GRBs at $z\geq6$.
Other high-$z$ ($\geq6$) GRBs include
GRB\,050904, GRB\,080913, GRB\,090423 and GRB\,140515A (e.g., Tagliaferri et al. 2005; Greiner et al. 2009; Tanvir et al. 2009; Melandri et al. 2015).

\it SVOM \rm is a China-France satellite mission dedicated to the detection and study of GRBs (Wei et al. 2016),
which is scheduled to be launched by the beginning of 2022.
The on-board payloads of \it SVOM \rm are able to detect GRBs in soft/hard $\gamma$-rays in near-real time,  and
to rapidly identify soft X-ray/optical counterparts or candidates in minutes. 
A on-ground VHF network evenly distributed on Earth under the satellite track is used to receive 
(refined) GRB's position and main characteristics in near-real time.
This capability combined with its anti-solar pointing strategy
allows the GRBs, especially high-$z$ ones, detected by \it SVOM \rm to be follow-up observed in spectroscopy
rapidly by ground-based large telescopes.

Rapid follow-up in spectroscopy of GRBs is in fact quite important in 1) discovery of high-$z$ GRBs through a redshift measurement; 2)  exploring structure and evolution of early universe through a study of reionization and 
cosmic star formation history;
and 3) examining the properties of the progenitors from the response of the circumburst gas to the central engine.
In this paper, we perform a pilot study on the spectroscopic follow-up in the \it SVOM \rm era by focusing on the
following two aspects: 1) the capability of rapid identification of optical candidates of high-$z$ GRB; and
2) an updated global photoionization response of circumburst gas to
GRB prompt emission (Wang et al. 2018) that is believed to be intensively addressed by the \it SVOM \rm mission.

The paper is organized as follows. Section 2 describes the conception and payloads of
the \it SVOM \rm mission in brief, by focusing on the detection of high-$z$ GRBs. The capability of \bf identifying \rm
optical candidates of high-$z$ GRBs is predicted in Section 3. Section 4 presents a
updated global photoionization response of GRB's local environment to the prompt emission.
A summary and perspective is shown in the last section.

\section{Profile and Instruments of \it SVOM}

We here briefly describe the profile and instruments of \it SVOM\rm, and
refer the readers to the White Paper given in Wei et al. (2016) for the details of the \it SVOM \rm mission.
The two wide-filed on-board instruments: the soft $\gamma$-ray imager ECLAIRs and the
Gamma-Ray Monitor (GRM) are designed to observe GRB prompt emission in 4-150 keV and 15-5000 keV
energy bands, respectively.  With a field-of-view (FoV) of 2sr and a sensitivity of $7.2\times10^{-10}\mathrm{erg\ s^{-1}\ cm^{-1}}$ (5$\sigma$ detection level in 1000 s), a total of 60-70 GRBs  per year
can be triggered by ECLAIRs, and 4-5\%\ of these GRBs are expected to have a redshift $\gtrsim5$
(Godet et al. 2014b; Cordier et al. 2015).

The two on-board narrow-field instruments: the Micro-channel X-ray Telescope (MXT, Gotz et al. 2014) 
and Visible Telescope (VT)
are responsible for follow-up observations of the afterglows in X-ray and optics, respectively.
VT is a Ritchey-Chretien telescope with a 44 cm diameter and a $f$-ratio of 9.
Its limiting magnitude is down to $m_V=22.5$ for a 300s exposure.
The FoV of VT is about $26\times26\mathrm{arcmin^2}$,  covering the ECLAIR's error box in most cases.
With a dichroic beam splitter, a GRB afterglow can be observed by VT in the two channels,
one in blue and the another in red,  simultaneously. The blue channel has a wavelength range from 0.4 to 0.65$\mu$m,
and the red one from  0.65 to 1.0$\mu$m. Figure 1 shows the normalized total throughput curves of the two channels,
along with the corresponding transmittance of the filters and the quantum efficiency (QE) of the
two $\mathrm{2k\times4k}$ E2V frame-transfer CCDs. A back-illuminated thick CCD is used for the
red channel to enhance the QE. More detailed description on the calibration and
determination of the throughput curves can be found in Qiu et al. (in preparation).

\begin{figure}
   \centering
   \includegraphics[width=8cm]{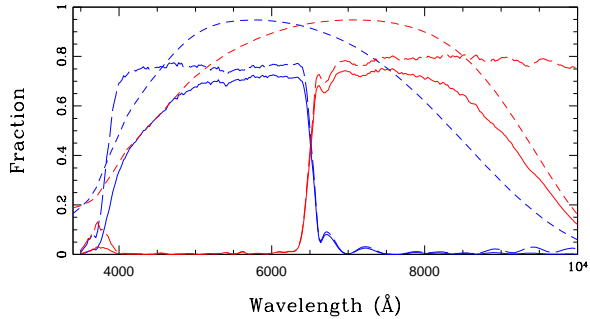}
   \caption{The normalized total throughput curves (the solid line), transmittance of the filters (the short-dashed line) and
QE of the CCDs (the long-dashed line) of VT as a function of wavelength. The blue and red channels are
denoted by blue and red colors, respectively.}
              \label{Fig1}%
\end{figure}

\section{Capability of Identifying of High- $z$ GRB Candidates: Color vs. Redshift}

In addition to high-$z$ quasars and galaxies (e.g., Fan et al. 2006; Bouwens et al. 2011a,b), the detection and follow-up spectroscopy of high-$z$ GRBs is a powerful tool in understanding the structure and evolution of high-$z$ universe.
On one hand, the short life-time of the massive progenitors  makes LGRBs a natural tracer of cosmic star formation
history (e.g., Vangioni et al. 2015; Wang et al. 2015; Petitjean et al. 2016), although some studies based on GRBs lead to an over-estimation
of star formation rate in the high-$z$ universe (e.g., Wang \& Dai 2009; Robertson \& Ellis 2012; Wang 2013a).
On the other hand, the spectroscopy of LGRBs with $z\geq6$ enables us to explore the reionization by ascertaining 
not only
the faint end slope of the galaxy luminosity function in the era of reionization but also the escape fraction of the 
ionizing radiation.
For instance, based on four $z\geq5.5$ LGRBs, Chornock et al.  (2014) reported some hints of a low value of  escape fraction in early universe.

The color provided by the simultaneous observations in the two channels in principle enables VT to provide a rapid identification of candidates of high-$z$ GRBs with a sub-arcsecond position accuracy, which is quite essential for 
subsequent rapid spectroscopy taken by ground large facilities. 
We here provide a quantified prediction on how to select high-$z$ GRB candidates based on the color.

\subsection{The VLT/X-shooter Spectroscopic Sample of GRB Afterglows}

We start our simulations with the X-shooter GRB afterglow legacy sample (XS-GRB, Selsing et al. 2019).
The sample in total contains spectra of 103 individual \it Swift \rm GRB afterglows that have been promptly observed with the ESO Very Large Telescope (VLT)/X-shooter cross-dispersed echelle spectrograph (Vernet et al. 2011) within 48 hours after the onset of the GRBs until 2017-03-31.  For a majority of the GRBs, the spectroscopy have been obtained by the 
X-shooter spectrograph by the UVB, VIS, and NIR-arms, which provides a spectral wavelength region from
3000\AA\ to 2.48$\mu$m. The spectral resolutions are 4350, 7450 and 5300 for the UVB, VIS and NIR-arm, respectively.

Because VT works in the wavelength range from 0.4 to 1.0$\mu$m, we select a sub-sample of GRBs from the XS-GRB sample by requiring the signal-to-noise ratio (S/N) of single pixel in the VIS arm is larger than 1.0. This selection results in
a sub-sample of 66 individual GRBs. The redshift of the sub-sample ranges from 0.156 to 5.91.
A spectral dataset with a wavelength range from 3500\AA\ to 1.1$\mu$m in the observer frame
are obtained for the 66 individual GRBs by a combination of the UVB, VIS, and NIR-arms.  
The combined spectra are  then smoothed by a boxcar of 10pixels to
enhance the S/N ratio. The telluric A-band (7600-7630\AA) and B-band (around 6860\AA) features due to $\mathrm{O_2}$ molecules are removed from the combined spectra by a linear interpolation.

\subsection{Color vs. Redshift}

With the pre-determined throughput curves of the two channels of VT, the corresponding two broad-band
AB magnitudes ($B_s$ and $R_s$ for the blue and red channels, respectively)
of the afterglow of each of the selected 66 GRBs are calculated according to the definition (Fukugita et al. 1996)
\begin{equation}
m_\mathrm{AB}=-2.5\log \frac{\int f_\nu S_\nu d\ln\nu}{\int S_\nu d\ln\nu}-48.6
\end{equation}
where $f_\nu$ is the specific flux density of the object in the unit of $\mathrm{erg\ s^{-1}\ cm^{-2}\ Hz^{-1}}$, and
$S_\nu$ the total throughput at frequency $\nu$.

The simulated colors $B_s-R_s$  are plotted against the spectroscopic redshifts in Figure 2 for the 66 GRBs.
One can learn from the
figure that 1) the $B_s-R_s$ color is insensitive to redshift for the burst with $z<3$. This is quite reasonable because
the Lyman break at $z\lesssim3$ is out of the wavelength coverage of the two channels of VT, 
and the $B_s-R_s$ color is mainly determined
by the powerlaw slope of the afterglow; 2) due to the gradually redshifted Lyman break, the
$B_s-R_s$ color increase from $\sim1$ to $\sim5$ steeply with $z$  when $z$ increases from $\sim3$ up to
$\sim6$. The $B_s-R_s$ vs. $z$ relationship can be parameterized by a three-order polynomial
$y=a_0+a_1z+a_2z^2+a_3z^3$, where $y=B_s-R_s$.  The best-fit curve is overplotted in Figure 2 by a long-dashed line,
and the corresponding values of $a_i\, (i=0-3)$ are listed in Table 1;
3) there are a few of low-$z$ outliers with large $B_s-R_s$ values ($B_s-R_s\sim1-2$). By examining their spectra
one-by-one by eyes, we find that the outliers are all resulted from their red spectra that is most likely due to the heavy local
extinction (e.g., Kruhler et al. 2011a; and references therein). Even though it is widely accepted that the dust up to $\sim30$pc away 
from the central engine within the beaming angle (e.g., Deng et al. 2010) can be completely destroyed by the high energy photons coming from the 
reverse shock (e.g., Waxman \& Draine 2000; Dai et al. 2001; Perna et al. 2003), the observed extinction
in the optical afterglow is possibly resulted from the dust outside of the initial beaming angle when the jet expands significantly (Liang et al. 2003).

Given our simulations based on the archival spectral database,  we can draw a conclusion that a color of $B_s-R_s>2$ can
be adopted as an effective way for selecting high-$z$ GRBs candidates with $z\sim4-6$,
while a contamination due to local dusty bursts must be taken into account when a bluer demarcation 
of  $B_s-R_s=1$ is adopted for selecting GRB candidates with $z\gtrsim3$.

\begin{figure}
   \centering
\includegraphics[width=8cm]{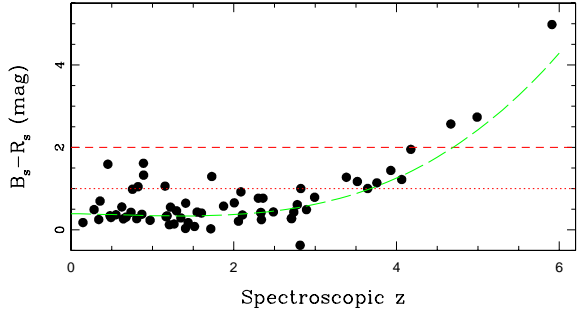}
\caption{A simulation based on the spectra of the 66 individual GRBs extracted from the XS-GRB legacy spectral database.
The $B_s-R_s$ color determined by the two VT channels is plotted against the spectroscopic redshift.  The best-fit three-order polynomial is overplotted by the long-dashed line.
The horizontal
dotted and short-dashed lines mark the proposed demarcation lines for $z\gtrsim3$ and $z\gtrsim4$ GRBs, respectively.
}
\end{figure}

\begin{table}
\renewcommand{\thetable}{\arabic{table}}
\centering
\caption{The best-fit values of $a_i\, (i=0-3)$}
\label{tab:decimal}
\begin{tabular}{cccc}
\hline\hline
$a_0$ &  $a_1$ & $a_2$ & $a_3$ \\
(1)  &   (2) & (3) & (4) \\
\hline
$0.392\pm0.107$  & $-0.030\pm0.163$  &  $-0.042\pm0.068$  &   $0.026\pm0.008$\\
\hline
\end{tabular}
\end{table}
   
\section{An Updated Global Photoionization Response to GRB Prompt Emission}

As aforementioned in the first section, ground-based spectroscopic follow-up observations of GRB afterglow can
greatly benefit from \it SVOM \rm both because of its capability of
rapid identification of optical candidates and because of its anti-solar pointing strategy.
In addition to determine the redshifts, the afterglow spectra of GRBs enable us to study the properties of both 
medium at GRB local environment and intervening absorption clouds located between the host galaxy and the
observer (e.g., Butler et al. 2003; Savaglio et al. 2003; Prochaska et al. 2006, 2007; Kawai et al. 2006; Totani et al.
2006; Thone et al. 2007; Tejos et al. 2007, 2009; Vreeswijk et al. 2007; D'Elia et al. 2009, 2010; Vergani et al. 2009; Wang 2013b; Kruhler et al. 2015).
The study of the circumburst environment can also in fact afford us some hint of GRB's progenitors.

Stemmed from some previous theoretical and observational studies and based on only 14 GRBs,
Wang et al. (2018 and references therein) proposed a global photoionization response of the
circumburst gas to GRB prompt emission, i.e., a relationship between the line ratio
CIV$\lambda$1549/CII$\lambda$1335 and either $L_{\mathrm{iso}}/E_{\mathrm{peak}}$ or
$L_{\mathrm{iso}}/E^2_{\mathrm{peak}}$. The CIV/CII line ratio traces the
ionization ratio of the interstellar medium (ISM) illuminated by a GRB, and
$L_{\mathrm{iso}}/E_{\mathrm{peak}}$  ($L_{\mathrm{iso}}/E^2_{\mathrm{peak}}$) the (specific) photon luminosity of the prompt emission.

In this section, we re-examine the CIV/CII-$L_{\mathrm{iso}}/E^2_{\mathrm{peak}}$ relation by doubling the size of
the previously used sample by supplementing the bursts extracted from the
XS-GRB spectroscopy database.
After acquiring reported values of both $L_{\mathrm{iso}}$ and $E_{\mathrm{peak}}$ from the literature,
there are  in total 14 bursts listed in the XS-GRB sample for subsequent spectral analysis.
We start our line profile modeling with a spectral smoothing with a boxcar of 10pixels to enhance S/N ratio,
since only equivalent widths (EWs) of the absorption features are needed in our study.
We then model each of the CII/CII$^*\lambda$1335 features by a single Gaussian absorption profile
by the SPECFIT (Kriss 1994) in
the IRAF package\footnote{IRAF is distributed by the National Optical Astronomical Observatories, which are
operated by the Association of Universities for Research in Astronomy, Inc., under cooperative agreement with the National Science Foundation.}.
Each CIV$\lambda\lambda$1548, 1550 doublet  is reproduced by two Gaussian functions, except for the two cases
GRB\,120119A and GRB\,1501301B, where only a total EW of the CIV$\lambda\lambda$1548, 1550 doublet can be obtained.
Columns (1) and (2) in Table 2 list the identification and the reported redshift for the 14 GRBs, respectively.
The measured EWs of  both CII/CII$^*\lambda$1335  feature and CIV$\lambda\lambda$1548, 1550 doublet in the
rest-frame are tabulated in Columns (3), (4) and (5) in the table. Columns (6) and (7) are the rest-frame
isotropic prompt luminosity ($L_{\mathrm{iso}}$) and peak energy ($E_{\mathrm{peak}}$) adopted from the
literature (Column (8)).
All the errors reported in the table correspond to the 1$\sigma$ significance level after taking into account
the proper error propagation. For the measured EWs, the uncertainties reported in Table 2 only include the statistical errors
resulting from the spectral fitting.

Figure 3 shows updated dependence of CIV/CII line ratio on 
both $L_{\mathrm{iso}}/E^2_{\mathrm{peak}}$ (right panel) and $L_{\mathrm{iso}}/E_{\mathrm{peak}}$ (middle panel) 
by combining the current XS-GRB sample and the sample previously used in Wang et al. (2018). 
The line ratio is also plotted against $L_{\mathrm{iso}}$ in the left panel in the same figure.
After excluding the five bursts with a upper limit of EW(CII$\lambda1335$), there are in total  28 bursts used
in the current study. One can see clearly from the figure there is no dependence of the line ratio on $L_{\mathrm{iso}}$.
Although there seems to be a  clear deviation at the large $L_{\mathrm{iso}}/E^2_{\mathrm{peak}}$ ($L_{\mathrm{iso}}/E_{\mathrm{peak}}$)
end, the dependence is confirmed and reinforced by the current study with enlarged sample for 
the bursts with $\log(L_{\mathrm{iso}}/E^2_{\mathrm{peak}})<47.5$ ($\log(L_{\mathrm{iso}}/E_{\mathrm{peak}})<50.4$), 
where $L_{\mathrm{iso}}$ and $E_{\mathrm{peak}}$ are in units of $\mathrm{erg\ s^{-1}}$ and keV, respectively.
We here argue that the deviation could be resulted from a wide distribution of the gamma-ray photons of the prompt emission,
although $E_{\mathrm{peak}}$ is a useful indication of the typical photon energy (e,g, Geng et al. 2018).


In the CIV/CII versus $L_{\mathrm{iso}}/E^2_{\mathrm{peak}}$ plot, 
after excluding the two outliers with extremely large CIV/CII line ratios, a non-parametric statistical test yields a Kendall's
$\tau=0.7179$ and a Z-value of 2.627 at a significance level with a probability of null correlation of
$P=0.0086$. We model the linear relationship as $y=\beta_0+\beta_1x+\epsilon$ according to
the D'Agostini fitting method (D'Agostini 2005, see also in, e.g., Guidorzi et al. 2006; Amati et al. 2008),
where $\epsilon$ is the extra Gaussian scatter. The updated optimal values obtained by the maximum likelihood method results in a relationship
\begin{equation}
 \log\frac{\mathrm{CIV}}{\mathrm{CII}}=(-8.01\pm1.41)+(0.17\pm0.03)\log\frac{L_{\mathrm{iso}}/\mathrm{erg\ s^{-1}}}{(E_{\mathrm{peak}}/\mathrm{keV})^2}
\end{equation}
with an extra scatter of $\epsilon=0.20$. The best-fitted relationship is overplotted in Figure 3.
Compared to our previous study in Wang et al. (2018), the current enlarged sample results in a much shallow dependence because of the deviation at the large $L_{\mathrm{iso}}/E^2_{\mathrm{peak}}$ end, although the 
intrinsic scatter $\epsilon$ is almost not changed. 

A poor correlation can be obtained for the dependence of the line ratio on $L_{\mathrm{iso}}/E_{\mathrm{peak}}$. 
The same statistics yields $\tau=0.4615$ and a Z-value of 1.689 at a significance level with a probability of null correlation of
$P=0.0913$ .

\begin{table}
\renewcommand{\thetable}{\arabic{table}}
\centering
\caption{A Sub-sample of XS-GRBs with Measured Absorption Lines and Prompt Emission}
\label{tab:decimal}
\footnotesize
\begin{tabular}{lccccccc}
\hline\hline
GRB &  $z$ & EW(CII/CII$^*\lambda1355$) & EW(CIV$\lambda1548$)    &  EW(CIV$\lambda1550$)    & $\log(L_{\mathrm{iso}})$ &  $\log(E_{\mathrm{peak}})$ & References  \\
        &         &  \AA & \AA  &  \AA  &   $\mathrm{erg\ s^{-1}}$ &  keV & \\
(1)  &   (2) & (3) & (4) & (5) & (6) & (7) & (8) \\
\hline
090926A &  2.106 &  $0.76\pm0.02$ & $0.51\pm0.01$ &  $0.41\pm0.01$  & $53.87\pm0.01$ &   $2.96\pm0.01$ & 1\\
111107A &  2.893 &  $1.99\pm0.05$ & $1.21\pm0.10$ &  $0.98\pm0.08$  & $52.24\pm0.08$ &   $2.62\pm0.13$ & 1\\
120119A$^a$ &  1.729 &  $4.31\pm0.07$ & $5.55\pm0.09$ &   \dotfill                & $52.90\pm0.06$ &   $2.62\pm0.06$ & 2 \\
120716A &  2.486 &  $2.18\pm0.03$ & $1.23\pm0.04$ &   $1.02\pm0.03$ & $52.67\pm0.14$ &   $2.79\pm0.22$ & 2\\
120815A &  2.359 &  $1.69\pm0.05$ & $1.20\pm0.06$ &   $0.89\pm0.05$ & $51.95\pm0.04$ &   $2.19\pm0.09$ & 1\\
130408A &  3.758 &  $1.50\pm0.04$ & $0.70\pm0.03$ &   $0.54\pm0.03$ & $53.76\pm0.05$ &   $3.00\pm0.06$ & 1\\
130612A &  2.006 &  $1.07\pm0.12$ & $0.61\pm0.24$ &   $1.08\pm0.25$ & $51.94\pm0.05$ &   $2.27\pm0.08$ & 1\\
141028A &  2.332 &  $0.91\pm0.01$ & $1.47\pm0.02$ &   $0.97\pm0.02$ & $53.27\pm0.04$ &   $2.92\pm0.02$ & 1\\
141109A &  2.993 &  $3.30\pm0.04$ & $2.90\pm0.10$ &   $2.26\pm0.10$ & $52.62\pm0.05$ &   $2.88\pm0.14$ & 1\\
150301B$^a$ &  1.517 &  $2.43\pm0.14$ & $2.40\pm0.06$ &   \dotfill                & $52.00\pm0.12$ &   $2.66\pm0.20$ & 1\\
150403A &  2.057 &  $2.73\pm0.02$ & $1.80\pm0.03$ &   $1.46\pm0.03$ & $53.60\pm0.05$ &   $3.06\pm0.08$ & 2\\
151021A &  2.333 &  $1.60\pm0.04$ & $1.24\pm0.06$ &   $0.82\pm0.06$ & $53.32\pm0.04$ &   $2.75\pm0.05$ & 1\\
160625B &  1.406 &  $1.68\pm0.05$ & $1.07\pm0.04$ &   $0.75\pm0.04$ & $54.07\pm0.04$ &   $3.14\pm0.02$ & 2\\
161023A &  2.710 &  $1.75\pm0.03$ & $0.87\pm0.03$ &   $0.73\pm0.03$ & $53.50\pm0.15$ &   $2.78\pm0.17$ & 2\\
\hline
\end{tabular}
\tablenotetext{a}{Only a total EW of CIV$\lambda\lambda$1548, 1550 doublets can be obtained.}
\note{References in the last column: (1) Ghirlanda et al. (2018); (2) Xue et al. (2019)}
\end{table}

\begin{figure}
\centering
\includegraphics[width=14cm]{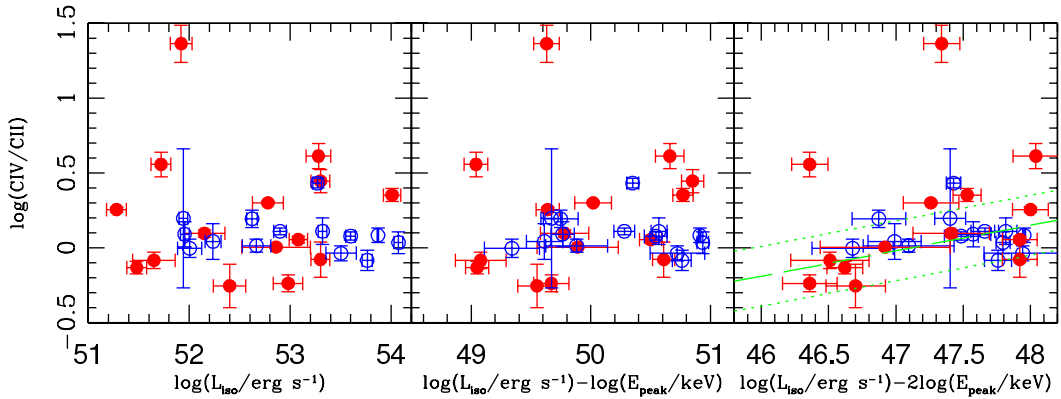}
\caption{An updated  CIV/CII-$L_{\mathrm{iso}}/E^2_{\mathrm{peak}}$ (right panel)  and  CIV/CII-$L_{\mathrm{iso}}/E_{\mathrm{peak}}$ 
(middle panel) sequences . The sample previously used in
Wang et al. (2018) is denoted by the red points and lines, and the added bursts measured in this work by the blue ones.
In the right panel, the best fit and 1$\sigma$ deviations obtained from  the D'Agostini fitting method (i.e., Eq. 2)  are plotted by the green 
long-dashed and dotted lines, respectively. \bf CIV/CII line ratio is plotted as a function of $L_{\mathrm{iso}}$ in the left panel, although there is no any dependence between the two variables.}
\end{figure}

\section{Summary}

In this paper,  a pilot study on the spectroscopic follow-up of GRB afterglows for the 
forthcoming \it SVOM \rm era is performed in two aspects.
On one hand, based on the archival XS-GRB spectral database, our simulation indicates that the $B_s-R_s$
color obtained from the SVOM/VT blue and red channels is effective for discriminating between  $z<3$ GRBs and
distant GRB candidates until $z\sim6$, although the latter are inevitably contaminated by the local dusty GRBs.
The contamination can be greatly reduced if we focus only on $z\gtrsim4$ GRBs.

On the other hand, by doubling the sample size through the XS-GRB spectral database,
the previously proposed global photoionization response of the circumburst gas to the prompt emission (i.e., the CIV/CII-$L_{\mathrm{iso}}/E^2_{\mathrm{peak}}$ relationship) is roughly confirmed, and
the confirmation is dissatisfactory at the large $L_{\mathrm{iso}}/E^2_{\mathrm{peak}}$ end. We state that 
thanks to  \it SVOM's\rm\ capability of rapid identification of optical candidates of afterglow and its anti-solar pointing strategy, a large spectroscopy sample is expected in the \it SVOM \rm era  to address this issue.


\begin{acknowledgements}
The authors thank the anonymous referee for his/her careful review and helpful suggestions that improved the manuscript.
The study is supported by the National Natural Science Foundation of China under grant 11773036 \& 11533003, and 
by the Strategic Pioneer Program on Space Science, Chinese Academy of Sciences, grant Nos. XDA15052600 \& XDA15016500.
JW is supported by Natural Science Foundation of Guangxi (2018GXNSFGA281007), and by Bagui Young Scholars Program.
   

\end{acknowledgements}

\label{lastpage}

\end{document}